\documentclass[twocolumn,dvipsnames]{aastex63}
\usepackage[utf8]{inputenc}
\usepackage{comment}

\begin{document}
\title[Finding High-Redshift Galaxies with JWST]{Finding High-Redshift Galaxies with JWST}


\correspondingauthor{Charles L. Steinhardt}
\email{steinhardt@nbi.ku.dk}

\author[0000-0003-3780-6801]{Charles L. Steinhardt}
\affiliation{Cosmic Dawn Center (DAWN)}
\affiliation{Niels Bohr Institute, University of Copenhagen, Lyngbyvej 2, DK-2100 Copenhagen \O }

\author{Christian Kragh Jespersen}
\affiliation{Cosmic Dawn Center (DAWN)}
\affiliation{Niels Bohr Institute, University of Copenhagen, Lyngbyvej 2, DK-2100 Copenhagen \O}

\author{Nora B. Linzer}
\affiliation{California Institute of Technology}
\affiliation{Cosmic Dawn Center (DAWN)}
\affiliation{Niels Bohr Institute, University of Copenhagen, Lyngbyvej 2, DK-2100 Copenhagen \O}

\begin{abstract}
One of the primary goals for the upcoming James Webb Space Telescope (JWST) is to observe the first galaxies. Predictions for planned and proposed surveys have typically focused on average galaxy counts, assuming a random distribution of galaxies across the observed field. The first and most massive galaxies, however, are expected to be tightly clustered, an effect known as cosmic variance.  We show that cosmic variance is likely to be the dominant contribution to uncertainty for high-redshift mass and luminosity functions, and that median high-redshift and high-mass galaxy counts for planned observations lie significantly below average counts.  Several different strategies are considered for improving our understanding of the first galaxies, including adding depth, area, and independent pointings.  Adding independent pointings is shown to be the most efficient both for discovering the single highest-redshift galaxy and also for constraining mass and luminosity functions.
\end{abstract}




\section{Introduction}
The past two decades have seen a rapid increase in our ability to detect, select, and spectroscopically confirm the existence of very high redshift galaxies.  Prior to the launch of the \emph{Hubble Space Telescope (HST)}, the highest-redshift known galaxies were detected because they were quasar hosts, and were limited to $z < 5$ \citep{Schneider1991}.  Following the discovery of the first $z > 5$ galaxy \citep{Dey1998}, new techniques have pushed the boundaries of our knowledge of galaxies out to $z = 11.1$ \citep{Oesch2016}. 

A primary goal of the next generation of telescopes is to find the very first galaxies to form in our observable Universe \citep{Rieke2019,Astro2010decadal}.  In particular, the \emph{James Webb Space Telescope (JWST)} has infrared cameras designed to allow the detection and spectroscopic followup of galaxies at $z > 11$ \citep{Gardner2006}.

However, these new capabilities provide no guarantee that it will be straightforward to actually find $z \gg 11$ galaxies.  Indeed, although \emph{HST} eventually discovered a galaxy at $z = 11.1$, it was not definitively confirmed until seven years after the most recent servicing mission.   The redshift record was held by a $z < 7$ source \citep{Iye2006} until 2009.  Finding these high-redshift galaxies required not only a technical improvement in our ability to observe them, but also new search techniques, both in the design of multi-wavelength surveys targeting the high-redshift Universe \citep[etc.]{Grogin2011,Koekemoer2011} and the novel technique of using gravitational lensing from massive clusters to exceed the typical \emph{HST} detection threshold \citep{Lotz2017,BUFFALO}.  

\emph{JWST} will similarly provide a major technical improvement in our ability to observe high-redshift galaxies over current telescopes.  However, \emph{JWST} is a time-limited resource, with a nominal mission lifetime of five years and a maximum telescope lifetime of $\sim 11$ years capped by consumables\footnote{https://earth.esa.int/web/eoportal/satellite-missions/j/jwst} and no servicing mission possible.  It is therefore essential to develop search strategies now, prior to launch, to ensure that \emph{JWST} will have time to both carry out these searches and conduct associated followup observations.  

Because first forays into the increasingly high-redshift Universe have historically yielded intriguing surprises, finding the optimal search strategy might only be possible after the first observations from \emph{JWST} have been completed.  Nevertheless, observational programs for the first period of operation have already been selected, based upon strategies which have been successful with \emph{HST}.  Programs targeting the earliest galaxies include the two JADES fields \citep{williams2018jwst}, modeled after the success of \emph{Hubble} deep fields, and GLASS, modeled after the success of a previous \emph{Hubble} program taking advantage of cluster lensing \citep{Treu2015}.

These programs were designed by extrapolating our knowledge of high-redshift galaxies to even higher redshift, then predicting the galaxy luminosity function and therefore the survey parameters required to find them.  This extrapolation is highly uncertain, and various models have extrapolated luminosity functions to conclude that \emph{JWST} will be able to find galaxies out to a maximum redshift of anywhere from $z = 13-16$ \citep{mashian2015empirical,williams2018jwst,jwsthighzpred,Behroozi2020}.   Although the single highest-redshift galaxy in a survey may be less meaningful than the broader distribution, it it commonly calculated for proposed surveys, in addition to providing a possible test of $\Lambda$CDM \citep{Behroozi2018}.  Thus, in this work, the single highest-redshift galaxy found is used as a figure of merit for comparing different survey strategies.

Each of these models and surveys has focused on how many high-redshift galaxies will be found in an \emph{average} case.  However, because the earliest galaxies must be formed in particularly rare, very overdense regions, high-redshift galaxies are not distributed randomly, but rather highly clustered, an effect known as cosmic variance (cf. \citet{Somerville2004}).  Cosmic variance increases both towards high redshift and towards high mass, and the earliest galaxies found by JWST will be the most massive galaxies at the highest redshifts ever observed.  Thus, JWST surveys will need to be designed accounting for stronger cosmic variance than in any previous survey in the history of astronomy.

The formalism and tools used here to extrapolate cosmic variance to high redshift are described in \S~\ref{sec:cv}.  Estimates of the high-redshift galaxy luminosity function are described in \S~\ref{sec:lumfunc}.  In \S~\ref{sec:surveys}, these are then applied to both currently planned or proposed surveys and several possible alternative survey designs.   The key conclusion is that cosmic variance is likely to dominate sample variance for most plausible surveys, and therefore additional coverage area and independent sightlines are more valuable than increased depth. Finally, the implications of these results are discussed further in \S~\ref{sec:discussion}.

Analysis presented here uses a flat $\Lambda$CDM cosmology with $(h, \Omega_m, \Omega_\Lambda) = (0.674 ,0.315, 0.685)$ \citep{Planck2018} throughout.  Magnitudes are given on the AB magnitude system \citep{Oke1974,Gunn1983}.

\section{Cosmic Variance}
\label{sec:cv}
Cosmic variance arises because the effects of large-scale structure influence the distribution of objects measured on smaller scales.  The variance between samples on small scales is therefore greater than the Poisson variance due to sample size alone.  This creates two significant problems when estimating cosmic variance.  First, there are no direct, confirmed observational measurements of large-scale structure even at the higher redshifts observed prior to JWST, let alone at $z > 10$.   There are several possible identifications of protoclusters \citep{Capak2011,Jiang2018,Hu2021}, which at these redshifts may still be in the process of expanding rather than contracting \citep{Muldrew2015}.

Second, our theoretical understanding of large-scale structure makes far most robust predictions about dark matter, since it comprises most of the mass, than it does about luminous galaxies.  This second problem is compounded by a lack of measurements of dark matter halos except in the very local Universe.  Thus, the connection between halo formation and galaxy formation cannot be reliably established at any nontrivial redshift.  

As a result, cosmic variance cannot be calculated directly at high redshift.  Rather, current estimates typically start from theoretical predictions of the halo mass function, then add additional assumptions about the relationship between galaxy luminosity and halo mass, which in practice is often two relationships, one between stellar mass and halo mass and a galactic mass-to-light ratio.  The most common of these, abundance matching, places the most luminous galaxies in the most massive halos, continuing from there towards lower luminosity and halo mass.  Because any two functions can be empirically matched, this assumption is inherently non-falsifiable even at redshifts where the galaxy luminosity function is well measured.  An alternative approach is instead to use empirical measurements of clustering to perform the matching, ensuring that the estimated cosmic variance will match observed variance at least at the redshifts where the matching is performed.  With any of these techniques, it will then be necessary to extrapolate the galaxy luminosity-halo mass relationship to even higher redshifts to estimate cosmic variance for JWST programs.

\subsection{Cosmic Variance Calculator}

\begin{figure}
\centering
\includegraphics[scale=0.35]{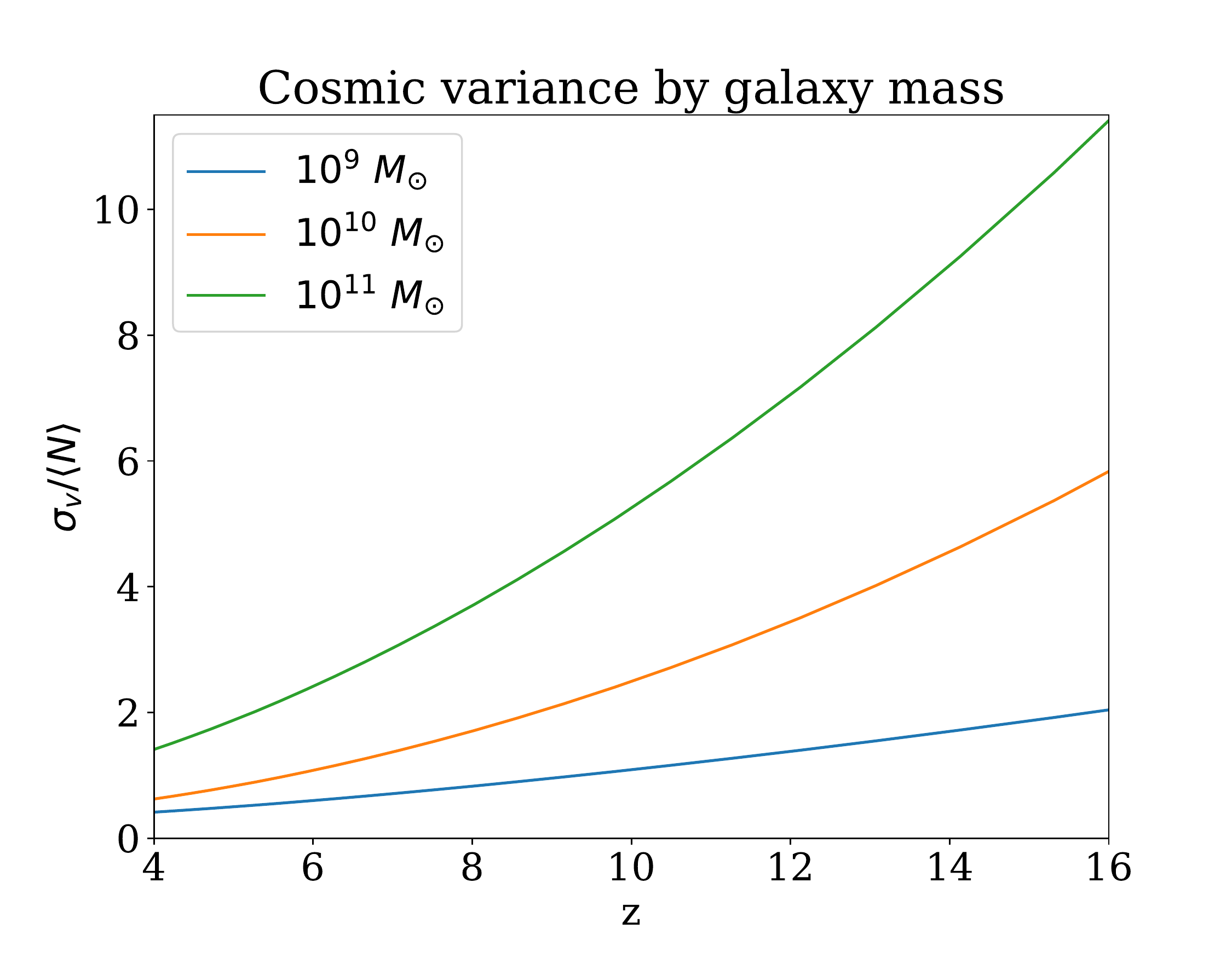}
\caption{Ratio of cosmic variance to the mean number of galaxies in a survey area of 10 arcmin by 20 arcmin as a function of redshift.  Cosmic variance grows towards high redshift and higher mass, and may be significantly greater than the mean for the first, most massive galaxies.}
\label{fig:cv}
\end{figure}

In this work, cosmic variance is estimated using the approach outlined in \citet{moster2011cosmic}.  Calculations are performed using a modified version of the \citet{moster2011cosmic} cosmic variance "cookbook" code, which has been updated to use cosmological parameters consistent with the \citet{Planck2018} results.  Cosmic variance is calculated in redshift bins that are equally spaced with  radial depth 150 Mpc in comoving distance.  The calculator assumes that each bin is independent from neighboring bins, so the width was chosen to approximate the scale of baryon acoustic oscillations \citep{Eisenstein2005,Alam2017}.  As an example, the results of the cosmic variance calculation at different redshifts and stellar masses for a survey of 10 arcmin by 20 arcmin are given in Figure \ref{fig:cv}.

 It should be noted that there are several more recent attempts to simulate or model cosmic variance at high redshift \citep{Trapp2020,Ucci2021}.  In particular, BlueTides simulation provides an updated variance calculator (cf. \citealt{Bhowmick2020}).  However, because BlueTides presents their results in H band, it cannot not be directly compared with this work.

\subsection{Resulting Probability Distribution}

Cosmic variance is often compared with the uncertainty due to sample size, which is often called either sample variance or Poisson variance.  The two terms are synonymous because in the absence of cosmic variance, the observed sample size follows a Poisson distribution with probability mass function (PMF; equivalent of the probability density function for a discrete distribution)
\begin{equation}
    \label{poisson}
    f(k;\lambda) = P(X=k) = \frac{\lambda^k e^{-\lambda}}{k!}.
\end{equation}
The expectation value of the Poisson distribution is $E(X) = \lambda$ and the variance is the familiar $\sigma^2 = \lambda,$ or $\sigma = \sqrt{\lambda} =  \sqrt{E(x)}$.  As $\lambda$ increases, the Poisson distribution approaches a Gaussian, with $\sigma/E(X) = 1/\sqrt{\lambda} \rightarrow 0.$

Cosmic variance results in a distribution with the same mean, but greater variance, than Poisson statistics would predict.  It is standard for variance calculators and simulations to report only the sample size mean and variance rather than the entire distribution.  However, for survey design the entire distribution is relevant.  For example,  the median cannot be calculated from mean and variance alone. 
\\
Here, we approximate the full probability distribution as a gamma distribution, 
\begin{equation}
    \label{gamma}
    f(x;k,\theta) = P(X=x) \frac{x^{k-1}e^{-\frac{x}{\theta}}}{\theta^k\Gamma(k)},.
\end{equation}
where $\Gamma$ is the Euler gamma function, an analytic continuation of the factorial with $\gamma(n) = (n-1)!, n \in \mathbb{N}.$  $k$ and $\theta$ are often called shape and scale, respectively.  The gamma distribution is in some sense an analytic continuation of the Poisson distribution, as can be seen from taking $\theta = 1$.

\begin{figure}
 \centering
 \includegraphics[scale=0.5]{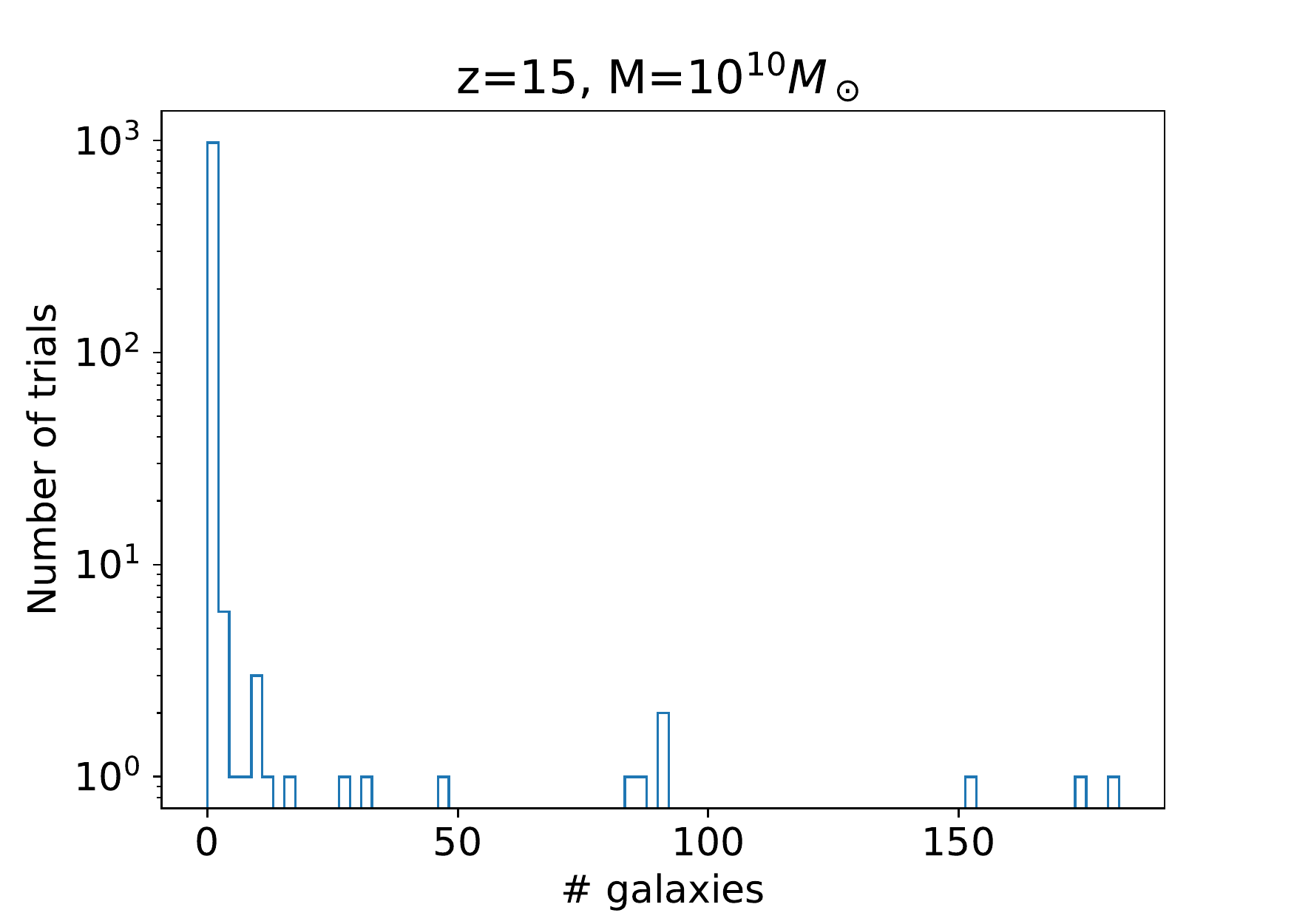}
 \caption{Result of running 1000 trials using a gamma distribution with a mean of 1 and variance given by the cosmic variance calculator for a JADES medium-like survey. In $\sim$ 96\% of the cases, no galaxies were seen.}
 \label{fig:trials}
 \end{figure}

The shape and scale can be used to separately set the mean ($E(X) = k\theta$) and variance ($\sigma^2 = k\theta^2$) of the distribution, taking:  
\begin{eqnarray}
\label{ktheta}
    k & = & \left(\frac{E[X]}{\sigma_v}\right)^2  \nonumber \\
    \theta & = & \frac{\sigma_v^2}{E[X]}. 
\end{eqnarray}
The entire distribution is then described by the mean, calculated using the extrapolated luminosity function as described in \S~\ref{sec:lumfunc} and the cosmic variance calculated here.  For small samples with cosmic variance significantly larger than the mean, the resulting distribution can be very highly peaked at 0, with the relatively rare fields containing at least one galaxy likely to contain many (Fig. \ref{fig:trials}).  The gamma distribution is compared with the distribution of sources in observed fields in \S~\ref{subsec:fieldvar}.

The gamma distribution is not the unique analytic continuation of the Poisson distribution.  Past work has also considered modeling rare events using a negative binomial distribution \citep{deSouza2015}.  This becomes ill-defined when $\mu \geq \sigma^2$,  where $\mu$ is the mean.  Here, the gamma distribution is selected, both because it produces valid results in all regimes and because it is observationally a reasonable approximation for observed distributions (\S~\ref{subsec:fieldvar}).

\section{Luminosity and Mass Functions}
\label{sec:lumfunc}

The full probability mass function depends not just on cosmic variance, but also on the mean and the sample variance.  As a result, modeling the population that will be observed by high-redshift surveys also requires extrapolating lower-redshift observations to determine the high-redshift galaxy distribution.  Further, because cosmic variance depends upon halo mass whereas observations depend upon luminosity, some technique for matching the two is required.  Here, we describe several possible techniques, as well as the reasons each is unlikely to be reliable at $z \gg 10$. 

\subsection{Luminosity Functions}
\label{subsec:lumfunc}
In principle, it might be best to work directly from observed luminosity functions (LF), extrapolating from high-redshift \emph{Hubble} surveys to estimate the $z>10$ LF.  \citet{bouwens2015uv} fit observed luminosity functions for $z \sim 4-10$ with a Schechter function, finding 
\begin{eqnarray}
\label{schechter}
\phi(M) & = & \phi^*\frac{\textrm{ln}(10)}{2.5}10^{-0.4(M-M^*)(\alpha+1)}e^{-10^{-0.4(M-M^*)}} \nonumber \\
    M^*_{UV} & = & (-20.95 \pm 0.10) + (0.01 \pm 0.06)(z-6) \nonumber \\
     \phi^* & = &  (0.47^{+0.11}_{-0.10})10^{(-0.27 \pm 0.05)(z-6)}10^{-3}\textrm{Mpc}^{-3} \\
     \alpha & = & (-1.87 \pm 0.05) + (-0.10 \pm 0.03)(z-6) \nonumber
\end{eqnarray}

\begin{figure}
\centering
\includegraphics[scale=0.4]{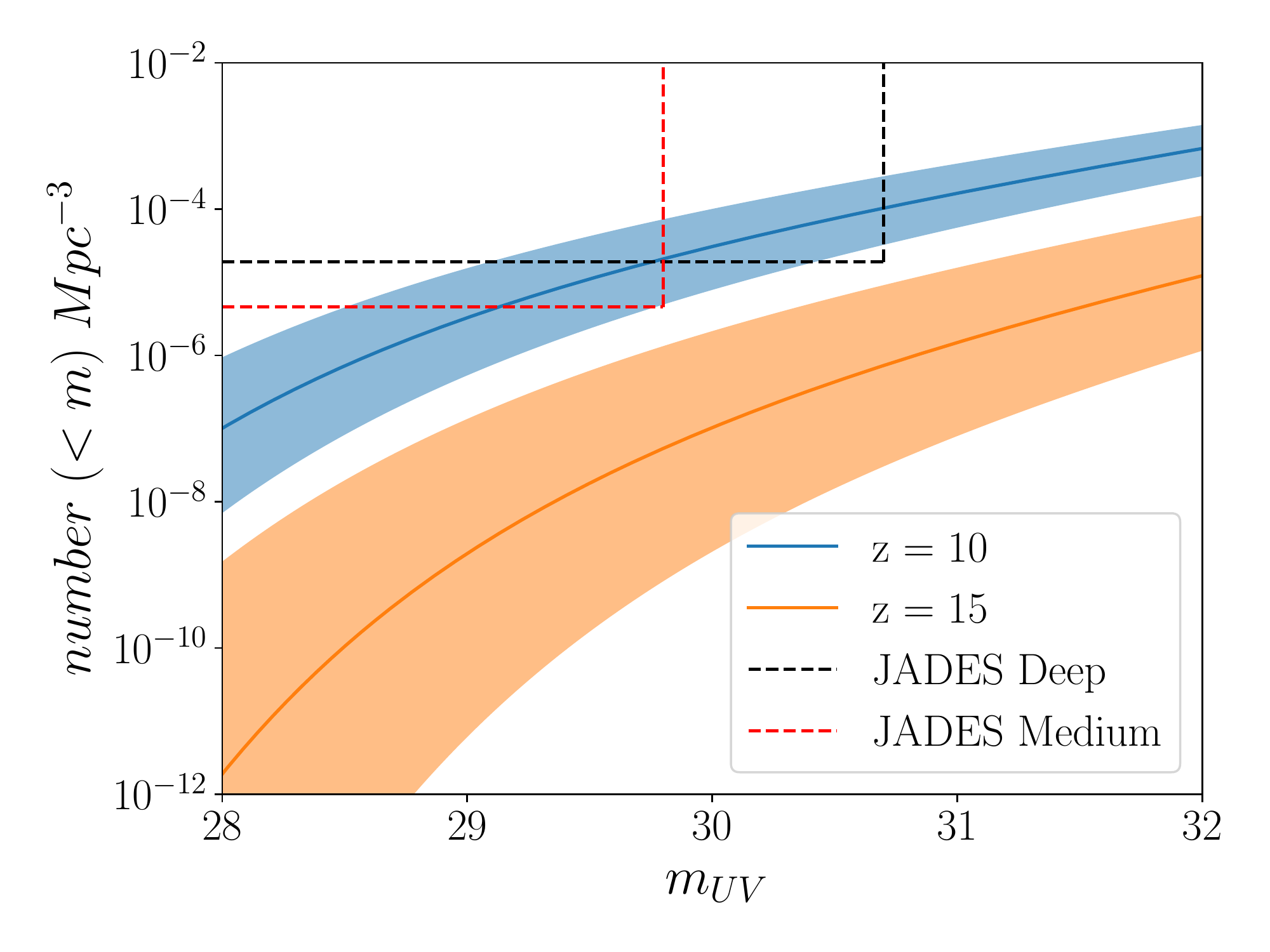}
\caption{Cumulative number density of galaxies as a function of apparent magnitude at redshift $10\pm 0.5$ and $15\pm 0.5$. Uncertainties shown reflect uncertainties in the \citet{bouwens2015uv} fit parameters, but no additional uncertainty due to extrapolation rather than interpolation.  The boxes indicate the expected coverage of the JADES medium (red, dashed) and deep (black, dashed) fields, respectively.}
\label{fig:lf}
\end{figure}

Using these fits, one can extrapolate the LF to $z > 10$. The uncertainties in the best-fit redshift dependence of $M^*$, $\phi^*$, and $\alpha$ lead to substantial uncertainties in the LF estimate at $z \gg 6$ (Fig. \ref{fig:lf}).  These results produce similar estimates to those reported by \citet{mashian2015empirical}, which concluded that an idealized JWST survey covering 200 $\textrm{arcmin}^2$ and with depth $\textrm{m}_{UV} = 31.5$ would on average be able to find a $z\sim 15$ galaxy.  The listed sensitivity of JWST/NIRCam in the exposure time corresponding to JADES Deep has a maximum limiting magnitude of 30.7 in the F200W-band, so this would correspond to an ultradeep survey.

Unfortunately, a luminosity function analysis alone is insufficient to determine the full PMF.  Cosmic variance is modeled as a function of halo mass, not luminosity, and therefore the two must be matched as well.  This cannot be done observationally, because except at very low redshift it has not yet been possible to determine halo masses directly.  Thus, the answer will instead depend upon the choice of model, which introduces an additional, substantial, and difficult to quantify uncertainty in estimating the PMF.

In keeping with the idea of trying to extrapolate as directly as possible from observations, one can convert from a UV luminosity function to a stellar mass function using the highest-redshift observations available.  \citet{song2016evolution} fit a linear relation between $\textrm{M}_{UV}$ and $\textrm{log}(\frac{\textrm{M}_*}{\textrm{M}_{\odot}})$. These best fit models are evaluated out to $z = 8$, and the fits at higher redshift are calculated by keeping slope fixed at the $z=8$.  This work adopts the same approach, using the best-fit $z=8$ slope at $z>8$ and extrapolating the \citet{song2016evolution} intercepts from a linear fit at $4<z<8$.  The end result is a mass-to-light ratio assumed to be universal at fixed redshift, which is then applied to convert between predictions of luminosity and stellar mass functions, as needed.  

Finally, these stellar masses must be converted to corresponding halo masses.  The stellar mass-halo mass relation cannot be observed directly outside of the local Universe, but \citet{moster2011cosmic} use an abundance matching-derived relation which is shown to yield a PMF consistent with observations out to $z \sim 3.5$.  

The uncertainties shown in Fig. \ref{fig:lf}, as well as the analysis in the remainder of this work, models uncertainties in the high-redshift LF based upon the fit uncertainties alone.  However extrapolation inherently assumes a strong similarity between the properties of high- and low-redshift galaxy distributions.  There are several strong reasons to expect that this may not be the case.  

The most luminous star-forming galaxies are more massive and more luminous towards high redshift \citep{Cowie1996,Glazebrook2004,Steinhardt2014,Somerville2015}, but this cannot continue indefinitely.  This has been greatly helpful for high-redshift astronomy, as the increased luminosity of the most luminous star-forming galaxies at high redshift somewhat mitigates the increase in luminosity distance.  However, this cannot go on indefinitely, since a galaxy cannot exist before its host halo has had time to form.  So, it should be fairly certain that the most massive star-forming galaxies at, e.g., $z=50$, will be less massive than those at $z=10$.  Any extrapolations from luminosity or mass functions beyond this transition will vastly overestimate the number of luminous sources.

Indeed, current observations suggest that this transition may lie below $z=15$.  It is possible that mass functions at $z<10$ are already difficult to reconcile with theoretical galaxy assembly under $\Lambda$CDM \citep{Steinhardt2016}; extrapolation to even higher redshifts would present an increasingly sharp problem (cf. \citet{Behroozi2018}, Fig. 2).  It is therefore natural to consider whether other models, although not directly rooted in observables, might be better predictors of $z \gg 10$ mass and luminosity functions.

\subsection{Halo Mass Function}
\label{subsec:hmf}

The most natural alternative approach is to begin with a halo mass function (HMF).  Here, HMFCalc \citep{murray2013hmfcalc} is used to generate the HMF as a function of redshift as well as its integral, the cumulative number density greater than some halo mass.  HMFCalc can be used with a variety of fitting functions, although the specific choice of model has negligible impact compared with the enormous uncertainties in the remainder of the problem.  As a result, even though the HMF is not observed directly, is constitutes a robust theoretical prediction.  For this work, calculations assume a best-fit Planck $\Lambda$CDM cosmology \citep{Planck2018} and a Sheth-Mo-Tormen fitting function \citep{Sheth2001} with $\Delta_\textrm{halo} = 200.0$ times the mean density.

Next, the halo mass must be converted to stellar mass.  The most common approach is to use abundance matching \citep{Kravtsov2004,Vale2006,Conroy2006} to determine the stellar mass-halo mass relation.  Abundance matching is inherently unfalsifiable, but is a reasonable approach if one is willing to assume that modeled halo mass functions are correct.  This assumption is not only appropriate but necessary here, because the remainder of the cosmic variance calculation already relies on this assumption.

Using abundance matching, \citet{finkelstein2015increasing} produce a best-fit stellar mass-halo mass relation, as well as its redshift dependence.  They also conclude that in order for this relation to hold, stellar baryon fraction must increase sharply at $z>4$.  This results in more massive and thus more luminous galaxies than would be produced from a local stellar mass-halo mass relation, the only one which can be determined observationally.  As with the LF, the abundance-matched stellar mass-halo mass relation is extrapolated to higher redshift for the calculations shown here.  

At sufficiently high redshift, this approach would even result in greater than 100\% of baryons residing in stars, which is almost certainly unphysical, but here the best-fit relation is used without correction.  Of course, if the high-redshift stellar initial mass function were sharply different (cf. \citet{Jermyn2018,Steinhardt2019}) or the stellar mass-halo mass relationship were different than produced by abundance matching, a smaller stellar baryon fraction could produce the same observations.

Finally, the relationship from \citet{song2016evolution} is again used, this time to convert from stellar mass to UV magnitude.  The result is then applied to predict the mean LF in a way that can be connected with halo mass, and thus cosmic variance. 

The additional assumptions made in this process introduce large systematic uncertainties.  The difference between a constant and evolving stellar baryon fraction alone produces a difference of more than 2 mag in the limiting depth required to find, on average, a redshift 15 galaxy (Fig. \ref{fig:lf}).  The difference between the local and abundance matched/extrapolated halo mass to stellar mass relation is even more significant by redshift 15.  Neither relation predicts the luminosity function extrapolated solely from observed luminosity functions, as in the previous section. 

In effect, the problem is that the relationships between halo mass and stellar mass, baryon mass, luminosity respectively, are nearly consistent with being constant out to $z \sim 4$, but then start to exhibit greater evolution at $z > 4$.  This makes extrapolation inherently difficult, and different choices for which relationships to use in that extrapolation will lead to different outcomes.  As a result, it is not possible to simultaneously maintain all of these lower-redshift relationships at high redshift, as they will yield inconsistent answers.  However, it should be possible to maintain whichever of these relationships is considered either best measured or theoretically most likely to extrapolate well out to high redshift, at the cost of breaking some of the other lower-redshift relations.  Indeed, many previous studies have done so \citep{Trenti2010,Tacchella2013,Dayal2014,Mason2015,Sun2016,Mirocha2017,Tacchella2018}.  For the purposes of this work, therefore, it is necessary to pick a relation as the basis for extrapolation.

\subsection{Choosing the Best Model}

Here we have discussed three approaches to modeling the very high-redshift galaxy distribution.  Each is a form of extrapolation, and relies on the assumption that the very high-redshift distribution can be well-modeled as an extension of lower-redshift distributions.  However, in each case there are good reasons to question whether that assumption will hold at $z \gg 10$:
\begin{itemize}
    \item {Extrapolating directly from observed luminosity functions yields the prediction which is most closely tied to observations. However, simulations at these redshifts suggest that the $z \gg 10$ Universe is a time primarily of galactic assembly rather than their subsequent evolution.  If so, extrapolating from properties of that evolution (e.g., the evolution of the star-forming main sequence \citealt{Speagle2014}) will produce a halo mass function inconsistent with simulations.}
    \item {Extrapolating instead from those simulations will produce a consistent halo mass function.  However, extrapolating the abundance-matched stellar mass-halo mass relation predicts that well over 100\% of baryons end up in stars by $z \sim 15$ \citep{Behroozi2018}.  This unphysical result should be an indication that such an extrapolation is invalid at such high redshifts.}
    \item {One could instead use a halo mass function from simulations, but a stellar mass-halo mass relation from the local Universe, the only place where it can be measured directly. This is guaranteed not to produce an unphysical result at high redshift.  However, it disagrees with the abundance-matched relation at $4 < z < 8$ \citep{finkelstein2015increasing}, implying that it is very unlikely to be valid at $z \sim 15$.}
\end{itemize}

Both luminosity function-based and halo mass function-based approaches agree out to $z \sim 7-8$, since they have been calibrated in that regime.  However, it is necessary to choose one of these approaches in order to discuss quantitative expectations from current and alternative survey designs.  In this work, we choose to extrapolate from luminosity functions to produce our primary results, as given in the shown figures.  This approach is most closely tied to observations, and historically theoretical predictions of new regimes in astronomy have a poor track record, which is part of why JWST observations are likely to be exciting.  It is also the approach taken most commonly in survey design whitepapers, making these results most easily comparable with predictions from other sources. Where possible, the predictions from multiple approaches are included to demonstrate that the general conclusions are independent of choice of analysis method.   

\subsection{Testing the Resulting Distribution}
\label{subsec:fieldvar}

The analysis here depends upon extrapolating cosmic variance calibrated at $z < 4$ in \citet{moster2011cosmic} to the highest redshifts observable with JWST.It will naturally be impossible to test the extrapolated distributions at $z \sim 12-15$ until the launch of JWST,  but some justification for the proposed use of the extrapolated cosmic variance can still be provided empirically, since newer observations in the COSMOS fields allow these extrapolations to be tested out to $z \sim 4-8$,  where the scenario $\sigma/\langle N\rangle > 1$ can be tested \footnote{$\langle N\rangle$ designates the average number of observed galaxies}.

The test is performed by splitting up the COSMOS field in a grid of JADES Deep - shaped (9.6' x 4.8') subfields, taking trials of each "pencil-beam" geometry in a standardized comoving volume  with radial depth 150 Mpc, centered at given redshifts. To avoid selection biases, we use the same flags as in \citet{davidzon17cosmos}, omitting fields with more than 10\% of the area missing. \\
 To perform a useful comparison of the galaxy counts in the COSMOS fields with the predicted gamma distribution, the analysis must be done in accordance with the cosmic variance cookbook, meaning that the analysis must be separated into the 150 Mpc radial depth bins, as well as binned by mass. Then, galaxy counts for every subfield at every redshift/mass bin are obtained. The mean of the galaxies across the subfields for the given redshift/mass bin is then computed, the variance for the given redshift and mass bin is obtained from  \citet{moster2011cosmic}, and the results are compared with a simulation following the gamma distribution with the same mean and the computed cosmic variance (Eqs. \ref{gamma}, \ref{ktheta}.)  The JADES Deep field is chosen to illustrate the applicability of the method in one of the smaller of the proposed survey areas, which suffer more from cosmic variance. If eq. \ref{gamma} models the JADES Deep field well, wider surveys should also be well modeled.

\begin{figure*}[!ht]
\centering
\includegraphics[scale=0.3]{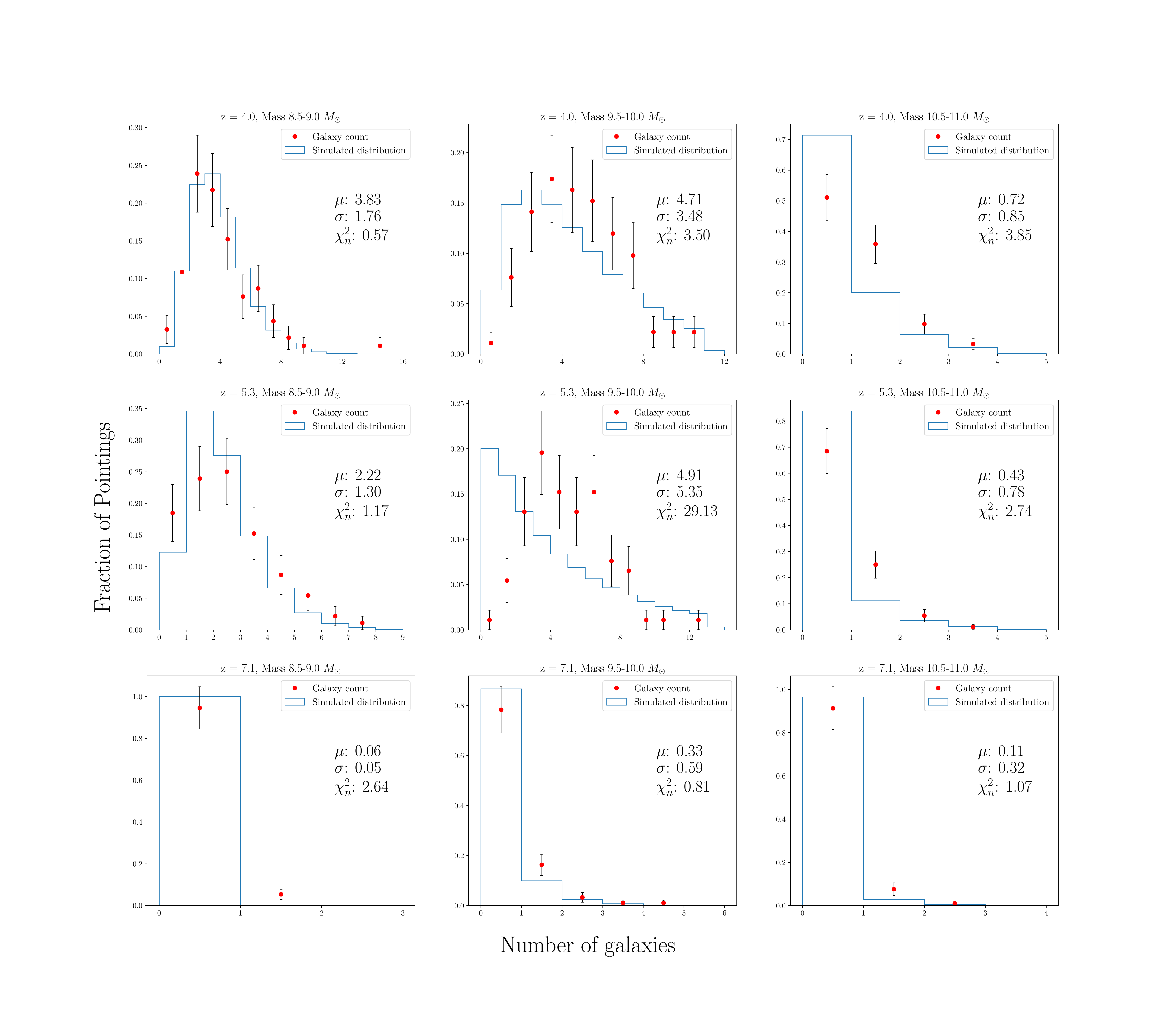}
\caption{Comparison between the predicted (blue) and observed (red) distribution of the number of COSMOS galaxies observed in independent 46 arcmin$^2$ pointings at $4.0 < z < 8.1$.  The gamma distribution produces a reasonable approximation for the distribution of sources between pointings for most cases when the overall number densities are matched, with the notable exception of the center panel ($z \sim 5.3, \log M_*/M_\odot \sim 9.75$).  Additional panels are shown in Fig. \ref{fig:allpanels}, at the end of the document.}
\label{fig:gammacomp}
\end{figure*}

Despite the extrapolations involved, the gamma distribution provides a reasonable approximation to the observed distributions (Figs. \ref{fig:gammacomp} and \ref{fig:allpanels}).  Where they differ, typically there are more observed regions with either very few or very many galaxies than in the gamma distribution, and therefore fewer in between.  This is seen most strongly in the center panel ($z \sim 5.3, \log M_*/M_\odot \sim 9.75$) of Fig. \ref{fig:gammacomp}, which appears to be an extreme outlier when compared against other COSMOS distributions at similar masses and redshifts.  It is unclear why this particular distribution is an outlier.  

The nature of differences between the gamma distribution and observations suggest that a gamma distribution typically underestimates the fraction of regions containing no observed galaxies (except when the mean is $\ll 1$).  As a result, this approximation should yield conservative results when considering the impact of cosmic variance on the median findings of JWST pointings.  The true impact will likely be larger than given in the following section.

\section{Survey Designs}
\label{sec:surveys}

Here, the implications of these results on survey design is explored.  To this point, whitepapers and proposals have nearly exclusively described \emph{average} quantities.  Here, we compare those with the \emph{median} results expected from these surveys.  As might be expected, the differences are minimal for common objects, but become stark when the most rare and extreme galaxies are considered.  As a common baseline, here different strategies are considered for discovering the single highest-redshift galaxy with JWST.  The following calculations are all done in comoving volumes with a radial depth of 150 Mpc. Comparisons for 100 Mpc and 250 Mpc radial depth can be found in tables \ref{tab:100Mpc},  \ref{tab:250Mpc}.  These comparisons do not provide much added information and are therefore found in the appendix.

\subsection{JADES}
Two fields which have already been selected for JWST observations are the JADES surveys.  One is a `wide' field, with a survey area of 190 $\textrm{arcmin}^2$ and limiting magnitude of 29.8 \citep{williams2018jwst}.  The `deep' field has an area of 46 $\textrm{arcmin}^2$, with a limiting magnitude of 30.7 in the F200W band.

\begin{figure*}
\centering
\includegraphics[scale=0.4]{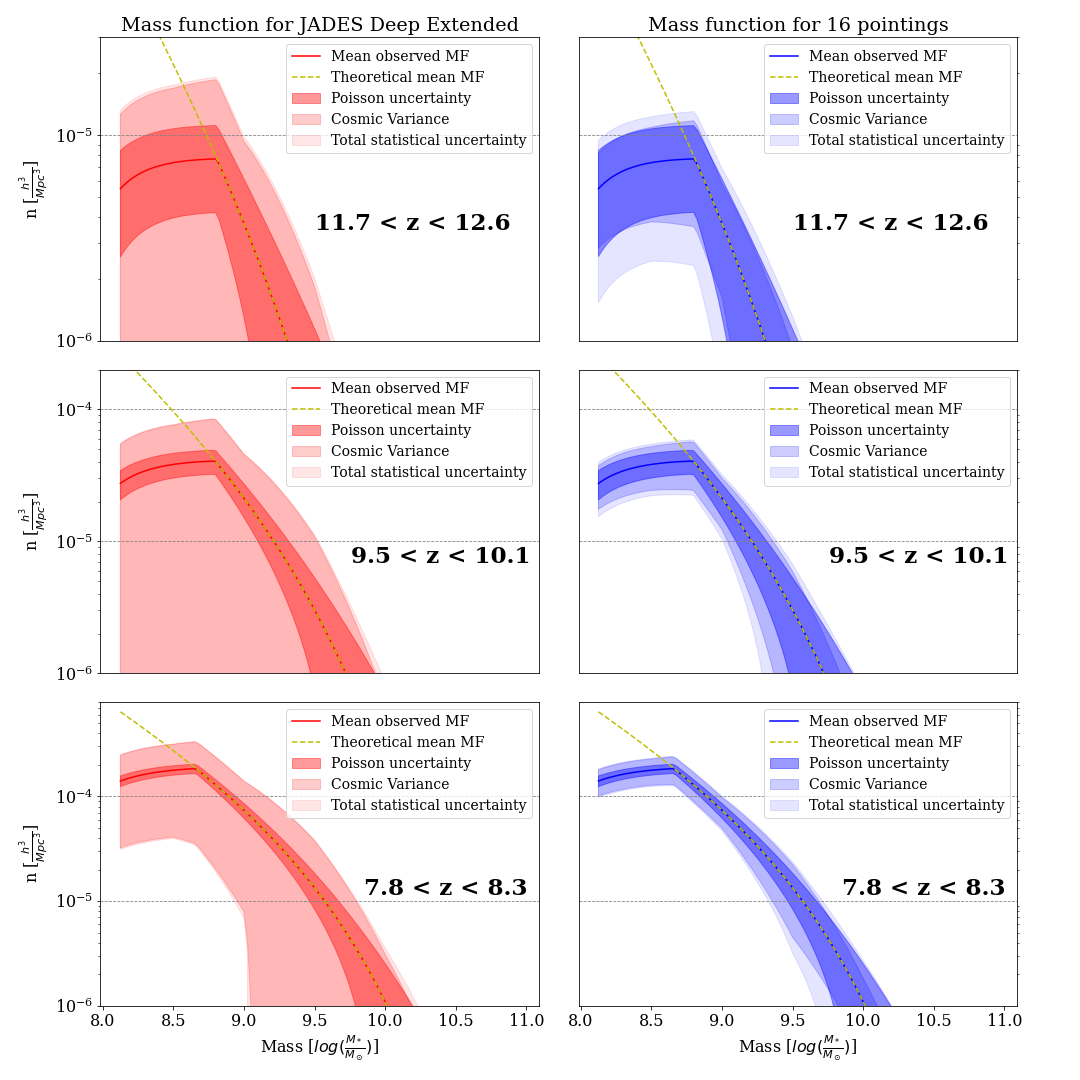}
\caption{A comparison of the theoretical mean number density of galaxies, mean observed number density, simulated uncertainties due to (Poisson) sample variance, cosmic variance, and a combination of the two (shaded lighter in that order) for an extension of JADES deep to a 1.0 Ms survey and a 16-pointing parallel survey of the same area and depth, which would also take 1.0 Ms.  At sufficiently high redshift or mass, the dominant source of uncertainty in a JWST mass or luminosity function will comes from cosmic variance, and the Poisson uncertainty contribution becomes negligible.  As a result, strategies designed to reduce cosmic variance by adding independent sightlines will be more effective than strategies designed to reduce sample variance by increasing the average number of high-mass, high-redshift galaxies found.  Surveys are compared over three different redshift ranges, with a total volume of $\sim 7 \times 10^5 \textrm{ Mpc}^3/h^3$ in each panel.}
\label{fig:jadeserrors}
\end{figure*}

Despite the different approaches, both perform similarly in searching for the highest-redshift galaxies.  With the standard assumptions described in \S~\ref{sec:lumfunc}, the JADES deep will, in the median case, find at least one galaxy out to $z = 13.1.$  As might be expected, at redshifts where many galaxies should be observed, the mean and median cases are similar.  However, at high redshift, where number counts become low, the highest-redshift galaxy in the median JADES-like survey lies at $z = 13.1$ and $z = 12.3$ for the deep and medium fields, whereas there is a mean of 1 galaxy at $z = 13.9$ and $z = 13.1$, respectively (Table \ref{tab:strategieslfext}).  This difference depends sharply on the extrapolation used.  A difference between the mean and median cases remains using any assumptions, but some extrapolations will produce a smaller difference (see Appendix) because they predict that the most luminous galaxies found by JWST will be more common. 

\begin{table}
\centering
{\bf Luminosity Function, Extrapolated SBF}
\begin{tabular}{|@{}|l|l|l|@{}|}
\hline
\textbf{Survey} & \textbf{Median=1}  & \textbf{Mean=1} \\
\hline
\textbf{JADES Deep} (0.31 Ms) & 13.1 & 13.9  \\
\textbf{JADES Medium} (0.24 Ms) & 12.3 & 13.1  \\
\textbf{JADES Deep (E)} (1.0 Ms) & 14.5 & 15.6  \\
\textbf{JADES Medium (E)} (1.0 Ms) & 14.2 & 14.9  \\
\textbf{1 pointing} (1.0 Ms) & 14.3 & 16.1 \\
\textbf{16 pointings} (1.0 Ms) & 14.8 & 15.6 \\
\textbf{20 pointings} (1.0 Ms) & 14.6 & 15.4 \\
\textbf{84 pointings} (1.0 Ms) & 14.3 & 14.9 \\
\textbf{100 pointings} (1.0 Ms) &  14.2 & 14.8 \\
\textbf{1000 pointings} (1.0 Ms) & 12.2 & 12.8 \\
\textbf{Lens 1$^\prime$x1$^\prime$} (63 ks) & 12.2 & 14.1 \\
   \hline
\end{tabular}
\caption{ Highest-redshift galaxy found in the median simulated survey using a variety of strategies, as well as the redshift at which there is a mean of 1 galaxy across similar surveys.  A variety of different strategies are compared, extrapolating luminosity functions to high redshift, then crossmatching the HMF with a linearly-extrapolated intercept.}
\label{tab:strategieslfext}
\end{table}

\subsection{Improving upon JADES}

Since a key goal for JWST is to find the earliest galaxies, it is natural to consider the best strategies for improving upon the JADES fields.  \citet{mashian2015empirical} propose an idealized ultradeep JWST survey which would have an area of 200 $\textrm{arcmin}^2$ and limiting magnitude of 31.5.  This would be a substantial program, requiring approximately 6.1 Ms.  However, the combination of increased depth and coverage area over JADES would have a substantial impact, as the median such survey would find a galaxy at $z = 17.5$.  The gap between the median and mean case also continues to increase, since these are even rarer, higher-redshift objects; the mean such survey would find a $z = 18.6$ galaxy.  Regardless, it is unlikely that such a survey will be completed in the early cycles of JWST.

A more realistic survey would be forced to choose between increased coverage and depth to limit the telescope time required.  Three strategies for doing so are considered here:
\begin{itemize}
    \item {{\bf Increased Depth:} Use additional time to increase the depth of the survey.  This has been the most successful strategy with \emph{HST}, taking advantage of gravitational lensing from foreground clusters to improve detection limits \citep{Treu2015,Lotz2017,BUFFALO}.  This would be more difficult at higher redshifts, since optimal magnification requires a massive cluster approximately halfway between the source and telescope and the most massive clusters at $z \sim 2$ are less massive than at $z \sim 0.5$.  However, a combination of lensing and additional observing time could be used to increase the depth of JWST surveys in small fields if a suitable lens could be found.  
      
    Here, a best-case scenario is considered in which lensing provides 2 magnitudes of improvement over a 1$^\prime \times 1^\prime$ field, with exposure time comparable to one JADES Deep pointing.     On average, this would find one $z = 14.1$ galaxy, appearing to provide an improvement over even the entire JADES Deep survey,  (Table \ref{tab:strategieslfext}).  However, the small coverage area suffers even more strongly from cosmic variance, and the median case ($z = 12.2$) performs poorly compared with JADES Medium and Deep.  Unless we are fortunate enough for one of the few $z~2$ protoclusters to line up perfectly with ultra-high redshift galaxies, the most successful strategy for \emph{HST} is unlikely to be the best strategy for \emph{JWST}.  Further, this approach cannot be scaled to wide surveys.
    \item {{\bf Broader Mosaic:} By expanding the JADES fields to a broader mosaic, there is an additional chance to find rare, high-redshift galaxies.  For example, given 1 Ms of JWST time, one could expand the JADES Medium field to a mosaic approximately five times the area.  The additional area would be contiguous with the existing JADES Medium field, and therefore galaxy counts would not be independent of the existing JADES fields.  As a result, there continues to be a significant difference between the mean ($z=15.1$) and median ($z=14.2$) cases.  So, there is a significantly greater chance of missing the first galaxies.  The main advantage is that this approach would allow exploration of larger-scale structures in these fields, but such structures are not predicted to exist at $z>10$,  despite the presence of a handful of possible protoclusters at somewhat lower redshifts \citep{Capak2011,Jiang2018,Hu2021}}.}
    \item {{\bf Many Independent Pointings:} A better strategy, then, would be to break the same 1 Ms up into many independent pointings.  As the number of pointings increases, the median and mean highest-redshift galaxy become increasingly similar (Table \ref{tab:strategieslfext}).  Perhaps the optimal combination in a dedicated survey would be  around 16 pointings of 62.5 ks exposure time, with a mean highest-redshift galaxy at $z = 15.6$ and median $z = 14.8$ or 20 pointings of 50 ks exposure time, with a mean highest-redshift galaxy at $z = 15.4$ and median $z = 14.6$}.  This strategy actually beats an expanded JADES Medium mosaic even though the expanded JADES field would also have a total exposure time of 1.0 Ms, demonstrating the importance of independent pointings.
\end{itemize}

In each of these scenarios, the most efficient strategies for finding the earliest galaxies with JWST require many independent sightlines.  Even taking advantage of previous JWST observations to make existing fields deeper or wider is likely to be less effective in the median case than observing entirely new fields.  Of course, once one of these rare, rich regions of early galaxy formation is found, the correct strategy will then change, and deeper observations around that initial detection will be optimal.

\section{Discussion}
\label{sec:discussion}

The ability of JWST to find high-redshift galaxies is primarily limited by its ability to find rare, rich regions of early galaxy formation more than by its ability to detect galaxies within those regions.  Because these are the most extremely overdense regions in our observable Universe, they also exhibit the largest cosmic variance.  As a result, there is a substantial difference between the mean and median number of sources surveys should expect to find.  For example, the model described in \S~\ref{sec:lumfunc} would predict that the average survey similar to JADES Medium will find two $z > 12.5$ galaxies, yet over half of all such surveys will find none. 

This effect is most pronounced for the most massive galaxies, where the average is small.  However, cosmic variance is expected to produce a significantly larger contribution to uncertainties than Poisson counting statistics for nearly the entire measured mass function at high redshift in a JADES-like survey (Fig. \ref{fig:jadeserrors}).  Thus, the optimal survey strategy must focus on reducing cosmic variance rather than increasing depth.

\subsection{Model Dependence of Optimal Strategy}

A significant concern is that the results presented here depend upon significant extrapolation of lower-redshift results.  Uncertainties shown in this work model uncertainties in the high-redshift LF based upon the fit uncertainties alone.   As discussed in \S~\ref{subsec:lumfunc}, there are several strong reasons to expect that this may not be the case. 

Regardless, in the absence of improved observational constraints, extrapolation cannot be avoided.  Further, cosmic variance will have similar qualitative effects regardless of the true luminosity function, because the limitations of any survey design will always lie near the detection threshold and for rare objects.  Thus, the primary lessons drawn here for JWST survey design should be expected to be applicable for any plausible high-redshift LF.  

Indeed, a comparison of several different extrapolations (Table \ref{tab:strategieslfext}; Appendix) finds that the redshift of the earliest galaxies found depends strongly on the technique.  However, the difference between mean and median cases, as well as the conclusion that adding independent pointings is the most efficient search strategy, is common to all of them.  The precise number of pointings and depths in that strategy, however, does depend upon the choice of extrapolation. 

The significant uncertainties in the number density and redshift of the highest-redshift galaxies found by JWST is a measure of the large improvement in capabilities that it will be provide.  Early observations with JWST will distinguish between these various models and point the way towards an optimal search strategy.  Even for those early observations, however, two key conclusions remain.  First, there is a significant difference between the mean and median outcomes, and survey design should focus on the median case.  And second, the most efficient way both to find the earliest galaxies and also to refine search strategies is to maximize the number of independent pointings rather than aiming for broader mosaics or deeper fields.

\subsection{Single-field Tests of \texorpdfstring{$\Lambda$CDM}{LCDM}}

Although the optimal strategy in most scenarios are similar, the details of what one should expect to find are model-dependent.  This makes the prediction problem more difficult prior to acquisition of JWST observations, but at the same time presents an opportunity to use those observations to distinguish between physical models for structure formation.  

The idea of using clustering as a test of the predictions of halo formation models is not novel.  It has already been possible in certain regimes to check for consistency by comparing two ways of matching luminous galaxies with halo masses, which depend upon predictions of $\Lambda$CDM in slightly different ways.  For a measured luminosity function and theoretical halo mass function, abundance matching (cf. \citet{Behroozi2019}) will produce a single solution for the relationship between luminosity and halo mass.  The success of this procedure does not provide a test of $\Lambda$CDM, as an incorrect halo mass function will also produce a single solution under abundance matching.  However, the results can be used to make a second set of predictions: if galaxies of a specific luminosity occupy halos of a known (abundance-matched) mass, then the clustering properties of those halos can be used to predict the clustering properties of galaxies at that luminosity. 

A clustering analysis of galaxies out to $z \sim 4$ produced clustering consistent with the abundance-matched halo masses at those redshifts \citep{Hildebrandt2009}.  At lower redshifts, two-point correlation functions have even been used successfully as tool for estimating the redshift distribution of a large galaxy sample \citep{McQuinn2013,Rahman2015,Morrison2017}.  However, these methods require good constraints on the two-point correlation function, which requires a large number of observed galaxies with well-understood completeness.

As a result, in practice it has been difficult to reduce the Poisson variance to the point that it might truly test $\Lambda$CDM.  Further, the ordering assumption made by abundance matching is very unlikely to be strictly true, since there is a non-negligible scatter in the halo mass-luminosity relation.   Because cosmic variance measurements rely on a small fraction of the total pairs in a sample, reassigning even a small fraction of objects can significantly change the resulting correlation function.  As a result, even if $\Lambda$CDM were incorrect, many test results could instead be interpreted as merely falsifying abundance matching.  An experiment which cannot falsify a model does not test that model; in a Bayesian framework, for example, the posterior probability will always equal the prior.

The new possibility for testing $\Lambda$CDM at high redshift comes from the dominant contribution of cosmic variance to observed galaxy counts.  As shown here, cosmic variance is often larger than Poisson variance at high redshift.  The effect is large enough to be measured reasonably well from even a small number of independent pointings.   Thus, if $\Lambda$CDM were sufficiently wrong, Poisson variance or reassigning objects to halos might not be able to explain the differences between prediction and measurement.  This test of $\Lambda$CDM would be another  study best performed on even a relatively shallow parallel survey.

Further, even the single highest-redshift source in one field is surprisingly sensitive to the halo mass-luminosity relationship.  Thus, potentially even the single highest-redshift galaxy found by JADES would hold the possibility of falsifying an incorrect halo mass function, and therefore allows a meaningful test of $\Lambda$CDM \citep{Behroozi2018}.  If indeed stellar mass functions even at $z < 10$ create tension with the predictions of standard cosmology, then perhaps even a single field with JWST will be enough to determine whether the problem is with the stellar masses or with the predictions $\Lambda$CDM makes for structure formation.

\acknowledgements
The authors would like to thank Gabe Brammer, Iary Davidzon, Pascal Oesch, Claudia  Scarlata, and John Weaver for useful discussions.  CLS is supported by ERC grant 648179 "ConTExt".  The Cosmic Dawn Center (DAWN) is funded by the Danish National Research Foundation under grant No. 140. 




\bibliographystyle{mnras}
\bibliography{references} 




\appendix

\section{Alternate extrapolations}

The results in Table \ref{tab:strategieslfext} extrapolate both observed luminosity functions towards high redshift and the variation in the stellar baryon fraction inferred from abundance matching at $z = 4-7$ \citet{finkelstein2015increasing}.  Here, the results are shown for two alternate extrapolations: (1) crossmatching the theoretical halo mass function with a linearly-extrapolated intercept rather than extrapolating luminosity functions; and (2) assuming that all galaxies share one constant $log(M/M_{\odot})-M_{UV}$ value rather than extrapolating the evolution of the inferred stellar baryon fraction.

The results in Tables \ref{tab:strategieslfcon}, \ref{tab:strategieshmfcon}, and \ref{tab:strategieshmfext} all produce the same qualitative results as in the main Table \ref{tab:strategieslfext}.  The median case is worse than the mean case, and a parallel survey with multiple independent pointings is the most efficient way to find high-redshift galaxies.  The details do change, and the highest-redshift galaxy expected in, e.g., JADES Deep, could range from $z=11-14$.  Further, the optimal combination of pointings and depth for an ideal parallel survey depends sharply on the assumptions made.  This is another way in which potentially the difference between independent JWST deep fields can be used as an additional test of which extrapolation is correct, and by extension, $\Lambda$CDM and assembly astrophysics.

\begin{table}
{\bf Luminosity Function, Constant M/L Ratio}
\centering
\begin{tabular}{|@{}|l|l|l|@{}|}
\hline
\textbf{Survey} & \textbf{Median=1}  & \textbf{Mean=1} \\
\hline
\textbf{JADES Deep} (0.31 Ms) & 13.1 & 13.9  \\
\textbf{JADES Medium} (0.24 Ms) & 12.2 & 13.1  \\
\textbf{JADES Deep (E)} (1.0 Ms) & 14.4 & 15.5  \\
\textbf{JADES Medium (E)} (1.0 Ms) & 14.2 & 14.8  \\
\textbf{1 pointing} (1.0 Ms) & 14.7 & 16.0 \\
\textbf{16 pointings} (1.0 Ms) & 15.1 & 15.5 \\
\textbf{20 pointings} (1.0 Ms) & 14.8 & 15.4 \\
\textbf{84 pointings} (1.0 Ms) & 14.3 & 14.8 \\
\textbf{100 pointings} (1.0 Ms) &  14.3 & 14.8 \\
\textbf{1000 pointings} (1.0 Ms) & 12.4 & 12.8 \\
\textbf{Lens 1'x1'} (63 ks) & 12.2 & 14.1 \\
   \hline
\end{tabular}
\caption{Highest-redshift galaxy found in the median simulated survey using a variety of strategies, as well as the redshift at which there is a mean of 1 galaxy across similar surveys.  A variety of different strategies are compared, extrapolating luminosity functions to high redshift while assuming that all galaxies at z $\geq$ 8 share one constant $log(M/M_{\odot})-M_{UV}$ value.}
\label{tab:strategieslfcon}
\end{table}

\begin{table}[ht]
\centering
{\bf HMF Matching, Constant M/L ratio}
\begin{tabular}{|@{}|l|l|l|@{}|}
\hline
\textbf{Survey} & \textbf{Median=1}  & \textbf{Mean=1} \\
\hline
\textbf{JADES Deep} (0.31 Ms) & 10.8 & 11.2  \\
\textbf{JADES Medium} (0.24 Ms) & 10.1 & 10.4  \\
\textbf{JADES Deep (E)} (1.0 Ms) & 11.4 & 11.7  \\
\textbf{JADES Medium (E)} (1.0 Ms) & 10.8 & 11.1  \\
\textbf{1 pointing} (1.0 Ms) & 12.3 & 13.0 \\
\textbf{16 pointings} (1.0 Ms) & 11.6 & 11.8 \\
\textbf{20 pointings} (1.0 Ms) & 11.5 & 11.7 \\
\textbf{84 pointings} (1.0 Ms) & 10.8 & 11.0 \\
\textbf{100 pointings} (1.0 Ms) &  10.7 & 10.9 \\
\textbf{1000 pointings} (1.0 Ms) & 9.6 & 9.9 \\
\textbf{Lens 1'x1'} (63 ks) & 11.9 & 12.4 \\
   \hline
\end{tabular}
\caption{Highest-redshift galaxy found in the median simulated survey using a variety of strategies, as well as the redshift at which there is a mean of 1 galaxy across similar surveys.  A variety of different strategies are compared, extrapolating luminosity and mass functions to high redshift by crossmatching the HMF while assuming that all galaxies at z $\geq$ 8 share one constant $log(M/M_{\odot})-M_{UV}$ value.}
\label{tab:strategieshmfcon}
\end{table}

\begin{table}[ht]
\centering
{\bf HMF Matching, Extrapolated SBF}
\begin{tabular}{|@{}|l|l|l|@{}|}
\hline
\textbf{Survey} & \textbf{Median=1}  & \textbf{Mean=1} \\
\hline
\textbf{JADES Deep} (0.31 Ms) & 11.3 & 11.7  \\
\textbf{JADES Medium} (0.24 Ms) & 10.4 & 10.8  \\
\textbf{JADES Deep (E)} (1.0 Ms) & 12.0 & 12.4  \\
\textbf{JADES Medium (E)} (1.0 Ms) & 11.0 & 11.5  \\
\textbf{1 pointing} (1.0 Ms) & 13.1 & 13.8 \\
\textbf{16 pointings} (1.0 Ms) & 11.9 & 12.4 \\
\textbf{20 pointings} (1.0 Ms) & 11.9 & 12.3 \\
\textbf{84 pointings} (1.0 Ms) & 10.9 & 11.4 \\
\textbf{100 pointings} (1.0 Ms) &  10.9 & 11.3 \\
\textbf{1000 pointings} (1.0 Ms) & 9.8 & 10.2 \\
\textbf{Lens 1'x1'} (63 ks) & 12.5 & 13.2 \\
   \hline
\end{tabular}
\caption{Highest-redshift galaxy found in the median simulated survey using a variety of strategies, as well as the redshift at which there is a mean of 1 galaxy across similar surveys.  A variety of different strategies are compared, extrapolating luminosity and mass functions to high redshift by crossmatching the HMF with a linearly-extrapolated $log(M/M_{\odot})-M_{UV}$ value.}
\label{tab:strategieshmfext}
\end{table}

Finally, an extended version of Fig \ref{fig:trials} is included here as Fig. \ref{fig:allpanels}.

\begin{figure*}[!ht]
\centering
\includegraphics[scale=0.275]{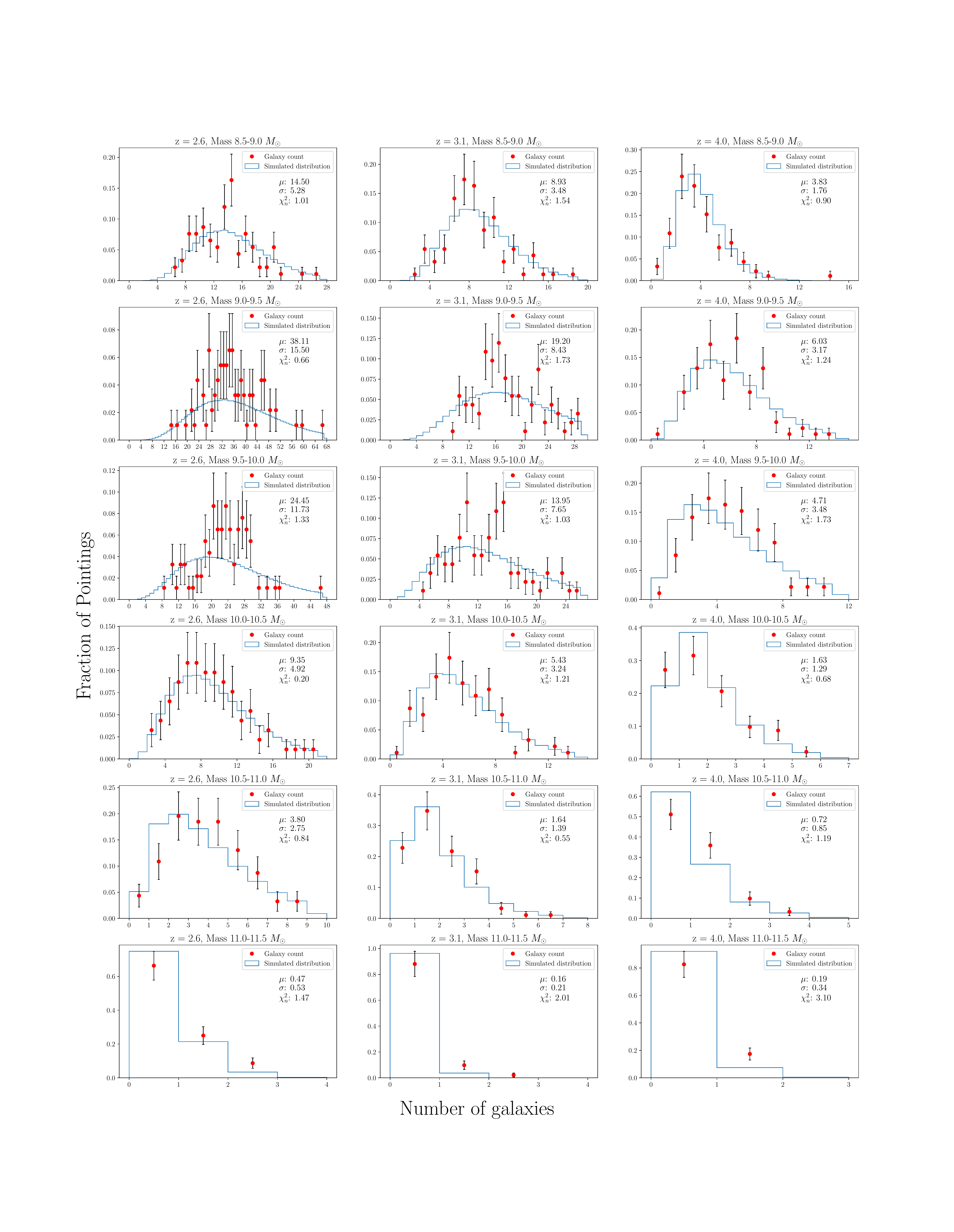}
\end{figure*}
\begin{figure*}[ht]
\includegraphics[scale=0.275]{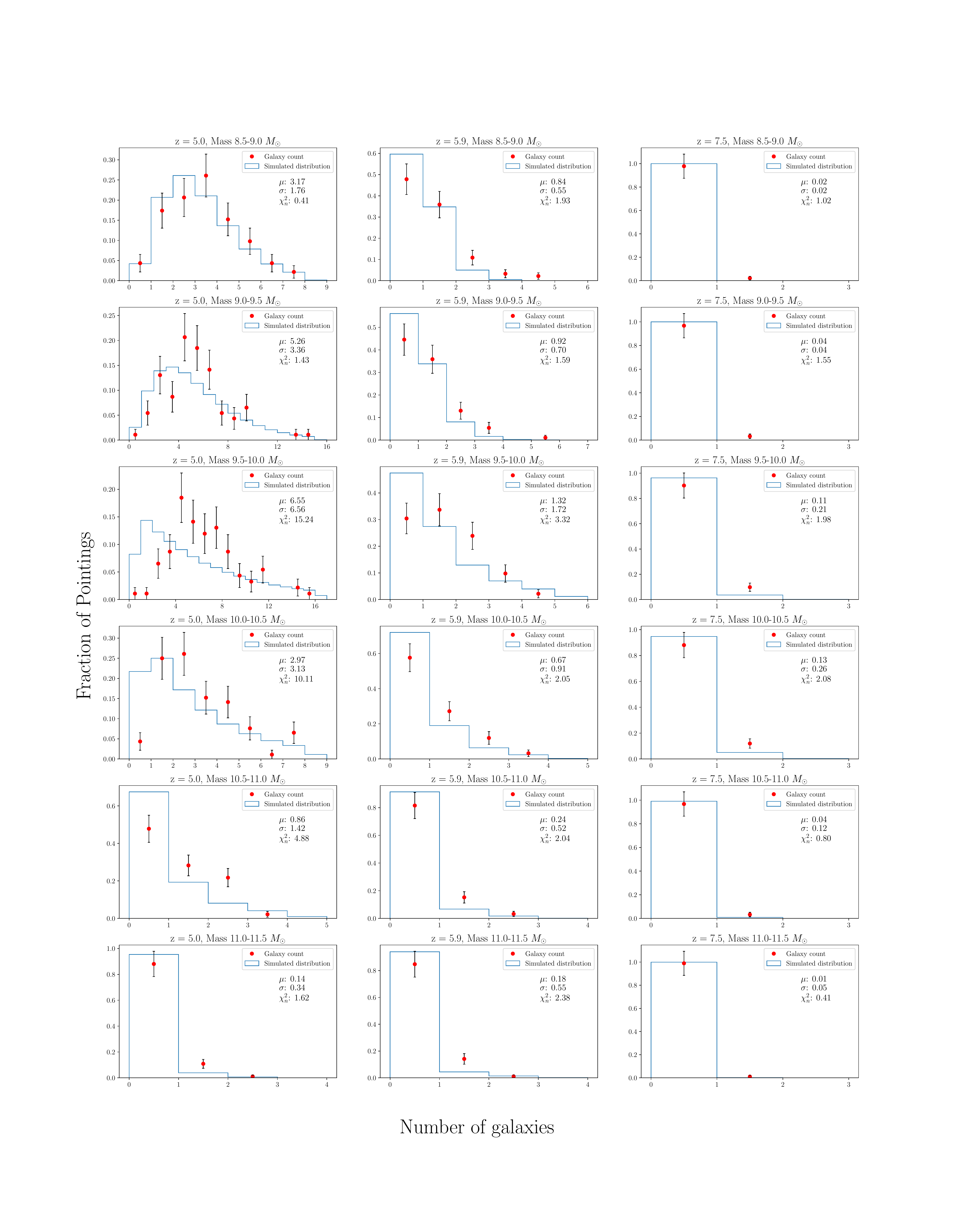}
\caption{(cont.) Comparison between the predicted (blue) and observed (red) distribution of the number of COSMOS galaxies observed in independent 46 arcmin$^2$ pointings at $4.7 < z < 8.1$.  The gamma distribution produces a reasonable approximation for the distribution of sources between pointings for most cases when the overall number densities are matched, with the notable exception of the center panel ($z \sim 5.3, \log M_*/M_\odot \sim 9.75$).}
\label{fig:allpanels}
\end{figure*}

\begin{table}
\centering
{\bf Luminosity Function, Extrapolated SBF (100 Mpc depth)}
\begin{tabular}{|@{}|l|l|l|@{}|}
\hline
\textbf{Survey} & \textbf{Median=1}  & \textbf{Mean=1} \\
\hline
\textbf{JADES Deep} (0.31 Ms) & 12.8 & 13.7  \\
\textbf{JADES Medium} (0.24 Ms) & 12.1 & 12.9  \\
\textbf{JADES Deep (E)} (1.0 Ms) & 14.5 & 15.4  \\
\textbf{JADES Medium (E)} (1.0 Ms) & 14.1 & 14.8  \\
\textbf{1 pointing} (1.0 Ms) & 14.2 & 15.8 \\
\textbf{16 pointings} (1.0 Ms) & 14.4 & 15.4 \\
\textbf{20 pointings} (1.0 Ms) & 14.3 & 15.3 \\
\textbf{84 pointings} (1.0 Ms) & 14.1 & 14.6 \\
\textbf{100 pointings} (1.0 Ms) &  14.0 & 14.5 \\
\textbf{1000 pointings} (1.0 Ms) & 11.7 & 12.6 \\
\textbf{Lens 1'x1'} (63 ks) & 12.1 & 14.0 \\
   \hline
\end{tabular}
\caption{ Highest-redshift galaxy found in the median simulated survey using a variety of strategies, as well as the redshift at which there is a mean of 1 galaxy across similar surveys. Here with a 100 Mpc radial depth.}
\label{tab:100Mpc}
\end{table}

\begin{table}
\centering
{\bf Luminosity Function, Extrapolated SBF (250 Mpc depth)}
\begin{tabular}{|@{}|l|l|l|@{}|}
\hline
\textbf{Survey} & \textbf{Median=1}  & \textbf{Mean=1} \\
\hline
\textbf{JADES Deep} (0.31 Ms) & 13.2 & 14.2  \\
\textbf{JADES Medium} (0.24 Ms) & 12.6 & 13.3  \\
\textbf{JADES Deep (E)} (1.0 Ms) & 14.8 & 15.9  \\
\textbf{JADES Medium (E)} (1.0 Ms) & 14.7 & 15.2  \\
\textbf{1 pointing} (1.0 Ms) & 14.7 & 16.4 \\
\textbf{16 pointings} (1.0 Ms) & 15.0 & 16.0 \\
\textbf{20 pointings} (1.0 Ms) & 14.9 & 15.9 \\
\textbf{84 pointings} (1.0 Ms) & 14.7 & 15.3 \\
\textbf{100 pointings} (1.0 Ms) &  14.6 & 15.1 \\
\textbf{1000 pointings} (1.0 Ms) & 12.6 & 13.1 \\
\textbf{Lens 1'x1'} (63 ks) & 12.8 & 14.3 \\
   \hline
\end{tabular}
\caption{ Highest-redshift galaxy found in the median simulated survey using a variety of strategies, as well as the redshift at which there is a mean of 1 galaxy across similar surveys. Here with a 250 Mpc radial depth.}
\label{tab:250Mpc}
\end{table}



\label{lastpage}
\end{document}